\documentclass[prl,reprint,amsmath,amssymb,superscriptaddress,longbibliography]{revtex4-2}

\usepackage{graphicx}
\usepackage{dcolumn}
\usepackage{bm}
\usepackage{hyperref}
\usepackage{breakurl}
\usepackage{multirow}
\usepackage{enumitem}
\usepackage{color}
\usepackage{gensymb}
\usepackage[utf8]{inputenc}
\usepackage[T1]{fontenc}
\usepackage{mathptmx}

\begin{document}

\title{Unlocking Hidden Spins in Centrosymmetric SnSe$_{2}$ by Vacancy-Controlled Spin-Orbit Scattering}

\author{Hengzhe Lu}
\thanks{Equal contributions}
\affiliation{Zhejiang Province Key Laboratory of Quantum Technology and Device, College of Physics, and State Key Laboratory of Silicon Materials, Zhejiang University, Hangzhou 310027, China}

\author{Zhibin Qi}
\thanks{Equal contributions}
\affiliation{Zhejiang Province Key Laboratory of Quantum Technology and Device, College of Physics, and State Key Laboratory of Silicon Materials, Zhejiang University, Hangzhou 310027, China}

\author{Yuqiang Huang}
\thanks{Equal contributions}
\affiliation{Zhejiang Province Key Laboratory of Quantum Technology and Device, College of Physics, and State Key Laboratory of Silicon Materials, Zhejiang University, Hangzhou 310027, China}

\author{Man Cheng}
\thanks{Equal contributions}
\affiliation{Zhejiang Province Key Laboratory of Quantum Technology and Device, College of Physics, and State Key Laboratory of Silicon Materials, Zhejiang University, Hangzhou 310027, China}

\author{Feng Sheng}
\affiliation{Zhejiang Province Key Laboratory of Quantum Technology and Device, College of Physics, and State Key Laboratory of Silicon Materials, Zhejiang University, Hangzhou 310027, China}

\author{Zhengkuan Deng}
\affiliation{Zhejiang Province Key Laboratory of Quantum Technology and Device, College of Physics, and State Key Laboratory of Silicon Materials, Zhejiang University, Hangzhou 310027, China}

\author{Shi Chen}
\affiliation{Institute of Applied Physics and Materials Engineering, University of Macau, Macao, China}

\author{Chenqiang Hua}
\affiliation{Beihang Hangzhou Innovation Institute Yuhang, Hangzhou 310023, China}

\author{Pimo He}
\affiliation{Zhejiang Province Key Laboratory of Quantum Technology and Device, College of Physics, and State Key Laboratory of Silicon Materials, Zhejiang University, Hangzhou 310027, China}

\author{Yunhao Lu}
\affiliation{Zhejiang Province Key Laboratory of Quantum Technology and Device, College of Physics, and State Key Laboratory of Silicon Materials, Zhejiang University, Hangzhou 310027, China}

\author{Yi Zheng}
\email{phyzhengyi@zju.edu.cn}
\affiliation{Zhejiang Province Key Laboratory of Quantum Technology and Device, College of Physics, and State Key Laboratory of Silicon Materials, Zhejiang University, Hangzhou 310027, China}
\affiliation{Collaborative Innovation Centre of Advanced Microstructures, Nanjing University, Nanjing 210093, China}

\date{\today}


\begin{abstract} 

Spin current generation and manipulation remain the key challenge of spintronics, in which relativistic spin-orbit coupling (SOC) play a ubiquitous role. In this letter, we demonstrate that hidden Rashba spins in the non-magnetic, centrosymmetric lattice of multilayer SnSe$_{2}$ can be efficiently activated by spin-orbit scattering introduced by Se vacancies. Via vacancy scattering, conduction electrons with hidden spin-momentum locked polarizations acquire out-of-plane magnetization components, which effectively break the chiral symmetry between the two Se sublattices of an SnSe$_{2}$ monolayer when electron spins start precession in the strong built-in Rashba SOC field. The resulting spin separations are manifested in quantum transport as vacancy concentration- and temperature-dependent crossovers from weak antilocalization (WAL) to weak localization (WL), with the distinctive spin relaxation mechanism of the Dyakonov-Perel type. Our study shows the great potential of two-dimensional systems with hidden-spin textures for spintronics.

\end{abstract} 


\maketitle


For a non-magnetic centrosymmetric bulk system, relativistic spin-orbit coupling effects are quenched due to the coexistence of both time reversal and inversion symmetries, which enforce spin double degeneracy for all electronic bands \cite{winkler_SOC_book,Rashba_SOC_review,Dresselhaus_PR_1955, rashba_1960_properties}. However, recent theoretical works reveal that many crystal systems host hidden spin polarization introduced by atomic-site inversion asymmetries, viz., local dipole filed (corresponding to R2-Rashba SOC) and site inversion asymmetry (D2-Dresselhaus SOC) respectively \cite{hidden_spin_origin_calculation}. Hidden spin textures, in particularly R2-Rashba, become an eminent feature in various van der Waals (vdW) crystals with the 1T-type transition-metal dichalcogenide (TMDC) monolayer structure, in which two opposite spin-momentum locked polarization directions within the vdW plane manifest the local dipole filed asymmetry exerted on the two inversion-symmetric chalcogen sublattices \cite{bulk_WSe2_hidden_spin, PtSe2_hidden_spin_Arpes}. Recently, many research efforts have been focused on main-group IV-VI dichalcogenides of GeS$_{2}$, SnS$_{2}$, and SnSe$_{2}$, aiming for diverse device applications ranging from thermoelectric \cite{thermoelectric}, optoelectronic \cite{photodetectors} to composite photo catalyzing \cite{photocatalyst}. Noticeably, the doubly degenerate electronic bands of these vdW semiconductors are also spin polarized by the innate strong R2-Rashba SOC, rooted in the 1T-TMDC monolayer lattice.

From spintronics point of view, these emergent 2D materials are promising spin transport channels for the next-generation spin logics and field-effect transistors \cite{FET_spin_transport, Origin_spintronics_review}, considering that the inherent SOC of these systems are order-of-magnitude larger than the intensively explored graphene \cite{graphene_spintronics}. However, spin transport and spin relaxation studies focusing on heavy-element 2D systems remain scarce at the moment \cite{graphene_spintronics, DP_MoS2_PRL}. Most critically, for hidden-spin 2D crystals, it seems to be formidable to achieve long spin diffusion lengths and long spin relaxation times, since there is a minimal energy cost for a conduction electron to hop between two chalcogen sublattices by flipping the spin polarization directions.

In this letter, we show that counterintuitively, spin transport and relaxation in the centrosymmetric lattice of multilayer SnSe$_{2}$ are not limited by random spin flipping, but rather tailored by \textit{motional narrowing} associated with the Dyakonov-Perel (DP) mechanism, in which the strong built-in R2-Rashba SOC function as the effective magnetic field for electron spin precession. The activation of the hidden Rashba spins are realized by spin-orbit scattering induced by Se vacancies, which enable conduction electrons to acquire out-of-plane magnetization components, and thus, effectively break the chiral symmetry between the two Se atomic layers. The resulting spin polarizations are manifested in quantum transport as vacancy concentration- and temperature-dependent crossovers from weak antilocalization (WAL) to weak localization (WL), which are excellently fitted by the Iordanskii-Lyanda-Geller-Pikus (ILP) theory. 

\begin{figure*}[htbp]
\includegraphics[width=7 in]{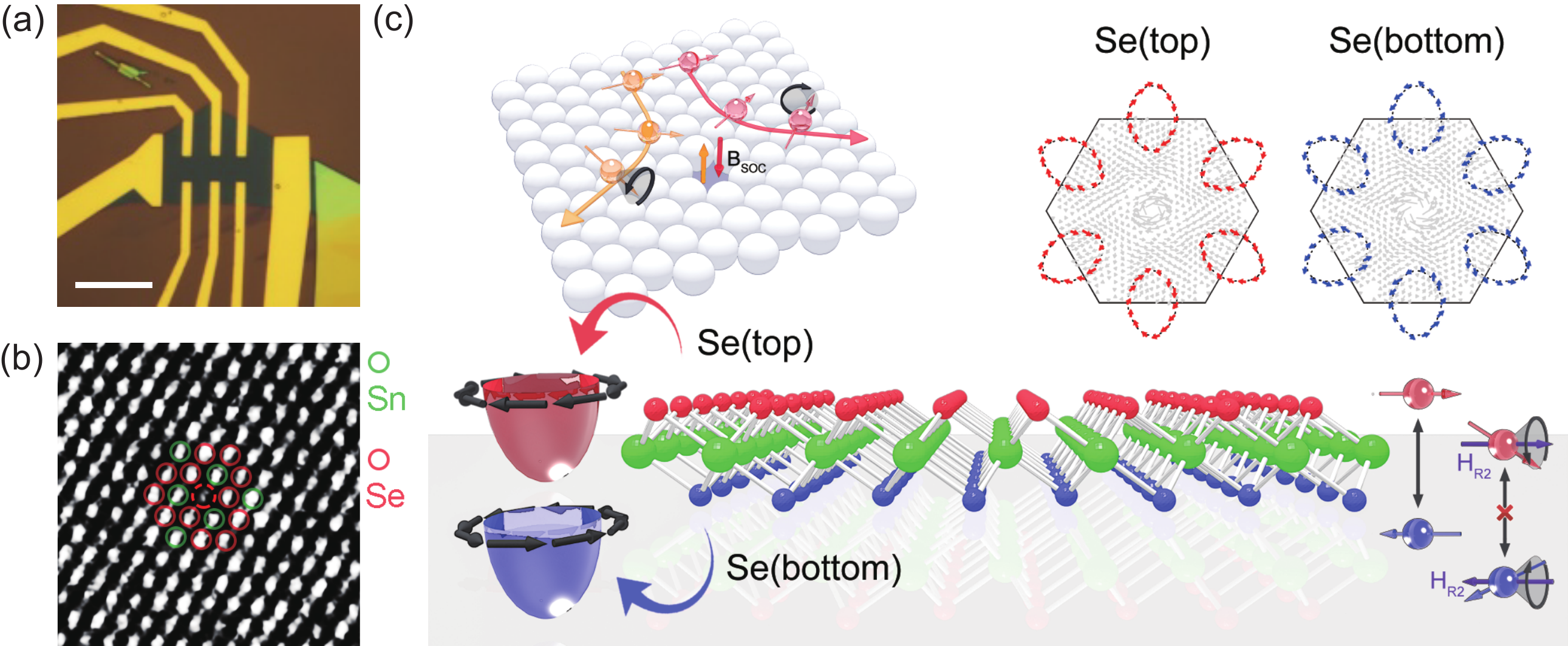}
\caption{Unlocking hidden Rashba spins in 1T-SnSe$_{2}$ by SOC scattering. (a) Typical optical image of a multilayer SnSe$_{2}$ device. Scale bar: 15 $\mu$m. (b) HRTEM image of few-layer SnSe$_{2}$, in which a single Se vacancy in darker contrast is highlighted by the red dashed circle. The trigonal lattice of  1T-SnSe$_{2}$ (space group ${P\overline{3}M1}$, No. 164) is apparent under HRTEM. (c) Vacancy-induced SOC scattering and the unlocking of hidden R2-Rashba spins  in 1T-SnSe$_{2}$. Upper left panel: SOC scattering by a negatively charged Se vacancy introduce effective SOC fields ($\mathbf{B}_\text{SOC}$) with opposite signs for left- and right-side scattered electrons respectively, i.e. the extrinsic spin Hall effect. Upper right: DFT calculated in-plane spin projection (grey arrows) of the first conduction band in the first Brillouin zone. The red and blue arrows are the hidden spin textures of a Fermi surface for $n=1\times10^{14}$ cm$^{-2}$. Note that experimental Fermi surfaces of $\sim10^{13}$ cm$^{-2}$ are too small for illustration. Bottom panel: sublattice-indexed chiral symmetry of 1T-SnSe$_{2}$. Spin processions break the chiral symmetry, and prohibit inter-sublattice hopping.}
\label{fig1}
\end{figure*}

SnSe$_{2}$ single crystals with controllable Se vacancy concentrations are synthesized by the self-flux method  \cite{wangzhen_SnSe_NC}. Note that despite the intermixed growth of SnSe and SnSe$_{2}$ \cite{wangzhen_SnSe_NC}, standard micro-mechanical exfoliations only yield thin SnSe$_{2}$ flakes, as shown in the Supplemental Material \footnote{See Supplemental Material at http://link.aps.org/supplemetal/ for Raman and AFM characterizations, DFT calculations of R2-Rashba hidden spins and electron doping by Se vacancies.} for a representative sample with consecutive thicknesses from monolayer to sextuple-layer. The absence of exfoliated SnSe flakes is likely due to the formation of unique interlayer point dislocations within the SnSe domains \cite{wangzhen_SnSe_NC}, since the elastic moduli of the two compounds are comparable. Nevertheless, all transport measured SnSe$_{2}$ devices (one representative optical image is shown in Figure \ref{fig1}(a)) have been further characterized by Raman spectroscopy to confirm the trigonal lattice structures, which have the fingerprinting $E_\text{g}$ and $A_\text{1g}$ phonon modes distinct from the puckering square lattice of SnSe \cite{wangzhen_SnSe_NC}. Using high-resolution transmission electron microscopy (HRTEM), we confirm the existence of randomly distributed Se single vacancies with a moderate density of $\sim10^{11}$ cm$^{-2}$ (Figure \ref{fig1}(b)), corresponding to electron doping of $n\sim10^{12}$ cm$^{-2}$ based on the density functional theory (DFT) calculations \cite{Note1}. Multilayer SnSe$_{2}$ field-effect transistors (FET) with typical channel thicknesses of $\sim 10$ nm for charge transport study were prepared by electron beam lithography followed by thermal evaporation of metal electrodes (5 nm Ti and 50 nm Au). The devices were then measured in a variable-temperature Oxford 14 T system using the standard four-probe technique with lock-in amplifiers.

As illustrated by the upper left of Figure \ref{fig1}(c), the presence of negatively charged Se vacancies \cite{Note1} cause strong SOC scattering for conduction electrons to acquire out-of-plane magnetization components, which is essentially the spin Hall effect (SHE) proposed by Dyakonov and Perel \cite{Dyakonov_Spin_Physics_in_Semiconductors}. Although similar extrinsic SHE phenomena have been reported in surface hydrogenated graphene \cite{H_graphene_SOC_Barbaros}, the inherent strong R2-Rashba SOC makes a fundamental difference in the spin relaxation mechanism of multilayer SnSe$_{2}$. Without vacancy-site SOC scattering, electron spins are in-plane polarized along two opposite spin-momentum locking directions with the same energy, as shown by the DFT calculated hidden R2-Rashba spin textures in the upper right panel of Figure \ref{fig1}(c) \cite{Note1}. Like pseudo-spins in graphene, the clockwise (counterclockwise) spin-momentum locking relation describes the instantaneous location of an electron spin wave function in the Se-top (Se-bottom) atomic layer of an SnSe$_{2}$ monolayer, which can be viewed as the sublattice-indexed chiral symmetry. Due to the energy degeneracy, conduction electrons are constantly hopping between two Se atomic layers by flipping the spin polarizations, suggesting long spin diffusion lengths and long spin relaxation times unrealistic for a hidden spin system like SnSe$_{2}$. However, once acquiring an out-of-plane magnetization, SOC scattered electron starts spin precession as driven by the strong built-in Rashba SOC field. As shown in the bottom panel of Figure \ref{fig1}(c), the spin procession effectively breaks the chiral symmetry between the two Se atomic layers, because inter-sublattice electron hopping now must enforce a reversal in the spin precession directions as required by the flipping of the built-in R2-Rashba SOC fields ($H_\text{R2}$).

\begin{figure}[htbp]
\includegraphics[width=3.5 in]{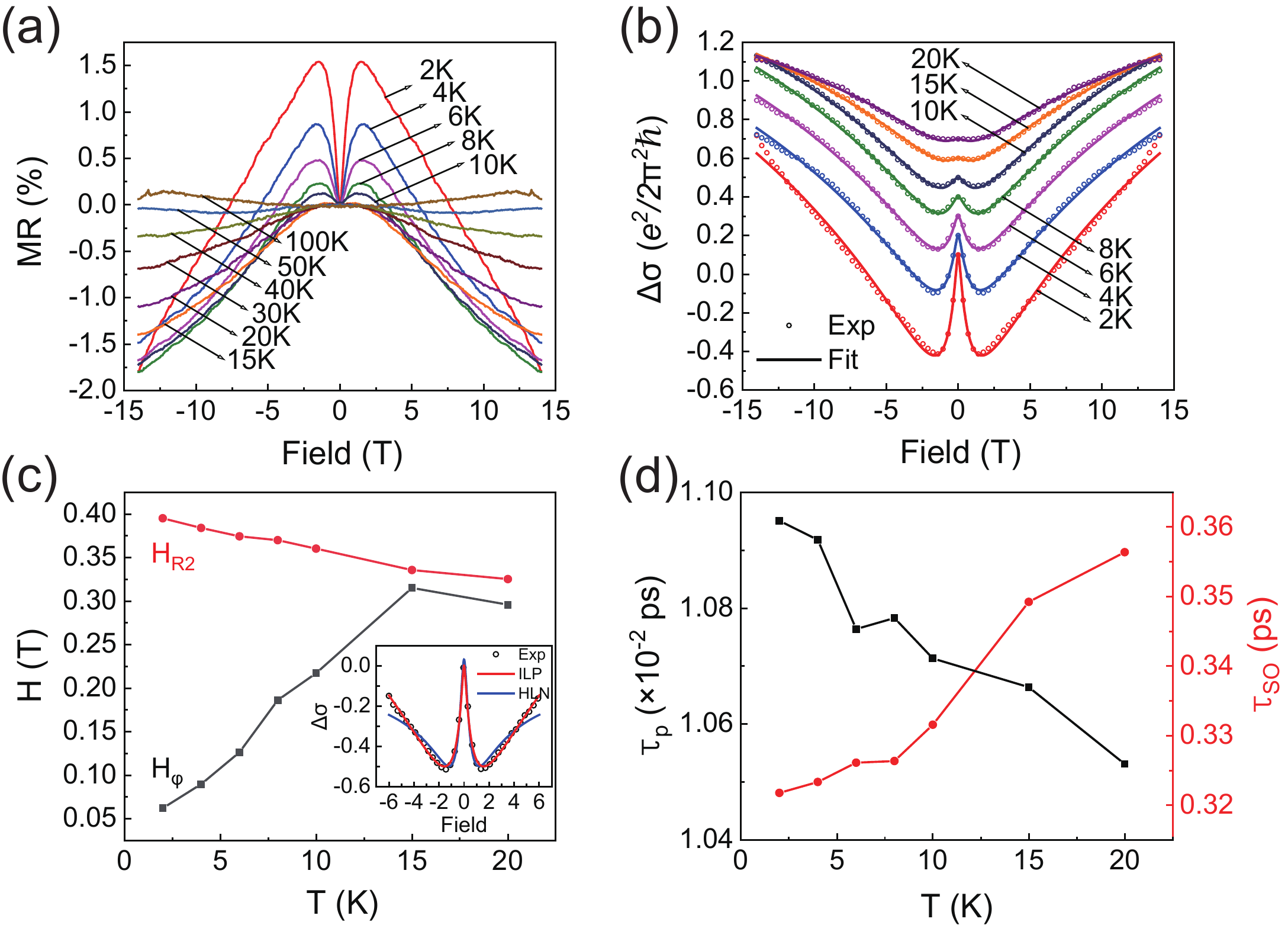}
\caption{Temperature-dependent WAL-to-WL transitions in multilayer SnSe$_{2}$. (a) $T$-dependent MR characteristics of SnSe$_{2}$ FET-SS5. Different MR curves are normalize by the zero-field minimum. (b) The corresponding MC curves, fitted by the ILP theory. Here, each MC curve is offset to show the remarkable consistency between the experimental results and the ILP model. (c) $H_{\varphi}$ and $H_\text{R2}$ extracted from the ILP theory fitting. The Inset compares the fitting results of the ILP theory with the HLN model. (d) $\tau_\text{SO}$ extracted from the ILP theory fitting, showing an inverse relation of $\tau_\text{SO} \propto \tau_\text{p}^{-1}$. }
\label{fig2}
\end{figure}

Intriguingly, by unlocking hidden Rashba spins using Se-vacancy scattering, spin polarizations obtain a finite relaxation time ($\tau_\text{SO}$) dependent on the competition between the spin precession time ($T_{1}\propto 1/H_\text{R2}$) and momentum scattering time ($\tau_\text{p}$), which can be quantitatively probed by temperature ($T$)-dependent quantum transport measurements \cite{wangyayu_PRL_WALWL_crossover, DP_MoS2_PRL, WZ_PRB_WAL, WAL_luli_prl}. As shown in Figure \ref{fig2}(a) for a representative device (SS5), helium temperature charge transport of few-layer SnSe$_{2}$ FETs are characterized by pronounced crossover behaviors from low-field WAL to high-field WL, hallmarking quantum interference corrections of spin polarized electron wave functions to the mesoscopic current flow \cite{WAL_calculation_luahizhou_1,WAL_calculation_luahizhou_2,wangyayu_PRL_WALWL_crossover}. By gradually warming up the sample, it is evident that the WAL phenomena are vanishing at 15 K, while the WL contributed negative magnetoresistance (MR) persist up to a much higher $T=40$ K. By plotting the data into the magnetoconductivity (MC) curves, we can clearly see the dwindling of the characteristic WAL cusps as a function of $T$, which can be fitted very well by the ILP theory \cite{ILP_HLN}, 
\begin{widetext}
\begin{equation}
\label{eqn:ILP}
\begin{split}
\Delta \sigma(B_{\bot}) = &\frac{e^{2}}{2\pi^{2}\hbar} \left\{ \Psi\left(\frac{1}{2}+\frac{H_{\varphi}}{B_{\bot}}+\frac{H_\text{R2}}{B_{\bot}}\right)\right. +\frac{1}{2}\Psi\left(\frac{1}{2}+\frac{H_{\varphi}}{B_{\bot}}+2\frac{H_\text{R2}}{B_{\bot}}\right)-\frac{1}{2}\Psi\left(\frac{1}{2}+\frac{H_{\varphi}}{B_{\bot}}\right) \\ &-\ln \left(\frac{H_{\varphi}}{B_{\bot}}+\frac{H_\text{R2}}{B_{\bot}}\right)-\frac{1}{2}\ln\left(\frac{H_{\varphi}}{B_{\bot}}+2\frac{H_\text{R2}}{B_{\bot}}\right)+\left.\frac{1}{2}\ln\left(\frac{H_{\varphi}}{B_{\bot}}\right)\right\},
\end{split}
\end{equation}
\end{widetext} 
in which ${\Psi}$ is the digamma function, $H_{\varphi}$ is the characteristic field of phase coherence \footnote {$H_{\varphi}$ is correlated to the phase coherence time $\tau_{\varphi}$ by the relation $\tau_{\varphi} = m^{*}/4\pi\hbar\mu nH_{\varphi}$. For quantum interference introduced conduction corrections, $H_{\varphi}$ is more intuitive than $\tau_{\varphi}$ for modeling and analyzing the experiment results, and the former is used for discussions in this letter.}, and $B_{\bot}$ represents the external magnetic field. As shown in Figure \ref{fig2}(b), the ILP formula not only reproduces the low-field WAL cusps but also captures the $B_{\bot}$-dependent WAL-to-WL crossover with remarkable consistency. In the seminal works of Pikus et al., it has been elucidated that the applicability of the ILP theory is a manifestation of the Dyakonov-Perel spin relaxation mechanism for the studied 2D systems \cite{DP_scattering, spin_procession_in_QWs}. In contrast, for an inversion symmetric system, WAL of spin degenerate electron wave functions should be activated by the Elliot-Yafet (EY) spin relaxation mechanism with $\tau_\text{SO} \propto \tau_\text{p}$, as modeled by the Hikami-Larkin-Nagaoka (HLN) theory \cite{EY_E_scattering, EY_Y_scattering, origin_HLN}. As compared in the inset of Figure \ref{fig2}(c), the HLN model can fairly fit the low-field WAL with an overshooting zero-field peak,  however, fails completely to conform the crossover regime and the succeeding WL growth. 

Using the ILP theory fitting, we are able to extract the $T$-dependent $H_{\varphi}$ and $H_\text{R2}$. As shown in Figure \ref{fig2}(c), below 15 K, $H_\text{R2}$ is much larger than $H_{\varphi}$, which explains the robust WAL behaviors in this $T$-regime. Noticeably, $H_\text{R2}$ exhibits a very weak $T$-dependence, which is expected for a build-in R2-Rashba SOC field. In contrast, $H_{\varphi}$ monotonically increases as a function of $T$, due to enhanced inelastic scattering at elevated $T$. The different $T$-dependence of $H_\text{R2}$ and $H_{\varphi}$ is responsible for the dwindling of the WAL cusps. Equally important, the ILP theory is microscopically originating in the DP spin relaxation mechanism, which correlates $\tau_\text{SO}$ inversely to $\tau_\text{p}$ by \textit{motional narrowing} \cite{Origin_spintronics_review, H_graphene_SOC_Barbaros}, i.e. frequent momentum scattering interrupt the full spin precession cycles and cause random walking in spin dephasing. For Se-vacancy scattered conduction electrons in SnSe$_{2}$, motional narrowing effectively increases the lifetime of the SOC induced out-of-plane magnetization, during which inter-Se-sublattice hopping costs finite energy to reverse the spin precession directions. By extracting $\tau_\text{p}$ using the Drude model \footnote{$m^{\ast}$ used here to calculate $\tau_\text{p}$ is deduced from the density functional theory calculations, because the Shubnikov-de Haas oscillations are too weak to extract convincing informations.}, we can indeed see the inverse relation of $\tau_\text{SO} \propto \tau_\text{p}^{-1}$, as shown in Figure \ref{fig2}(d).

\begin{figure*}[htbp]
\includegraphics[width=7 in]{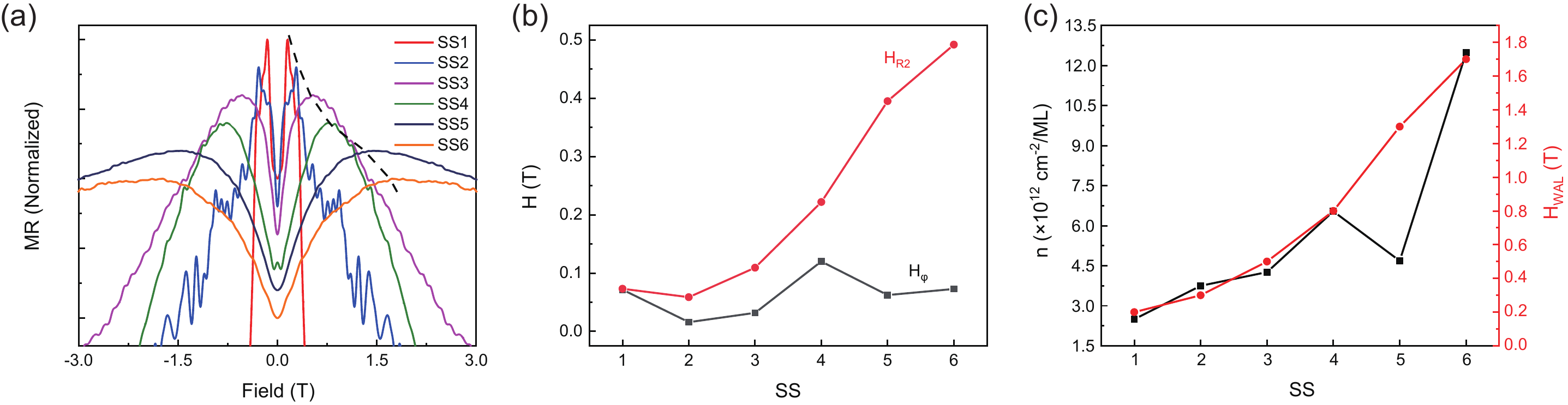}
\caption{WAL-to-WL crossovers vs Se-vacancy concentrations in SnSe$_{2}$. (a) MR curves of six multilayer SnSe$_{2}$ FETs (designated as SS1-SS6 respectively) with increasing electron doping, which is proportional to the Se-vacancy density. Different curves, normalize by the zero-field minimum, are offset for clarity. The black dashed line indicates the positions of the MR maximums (H$_\text{WAL}$) for different devices. (b) $H_{\varphi}$ and $H_\text{SO}$ extracted from the ILP theory fitting. (c) H$_\text{WAL}$ extracted from the MR curves in Figure 3(a), showing the same growth trend of $H_\text{R2}$ when Se-vacancy concentrations increase.}
\label{fig3}
\end{figure*}

The critical role of Se-vacancy scattering for the activation of hidden Rashba spins is also demonstrated by a positive dependence of $H_\text{R2}$ on electron doping levels ($n$), which are directly controlled by Se vacancy densities. As shown in Figure \ref{fig3} for six different FETs (designated as SS1-SS6) measured at $T=2$ K, with the presence of more Se vacancies, conduction electrons have increased SOC scattering rates to acquire out-of-plane magnetization components, and thus, $H_\text{R2}$ monotonically grows while $H_{\varphi}$ remains nearly unchanged. By defining the WAL maximum in different MR curve as $H_\text{WAL}$, which also reflects the magnitude of $H_\text{R2}$, it is also clear that increased Se-vacancy scattering contributes positively to spin polarizations and $\tau_\text{SO}$ in SnSe$_{2}$. 

\begin{figure}[htbp]
\includegraphics[width=3.5 in]{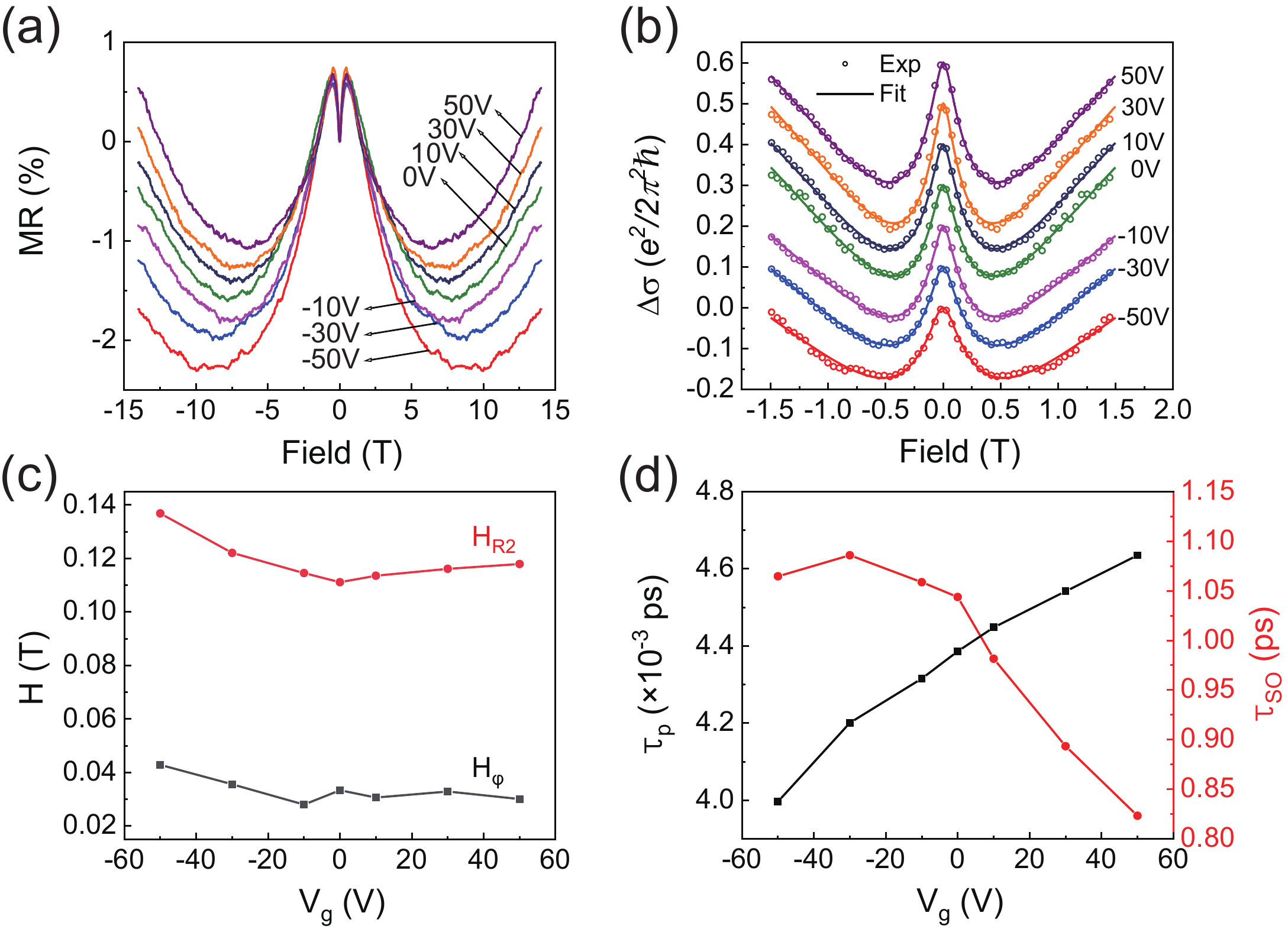}
\caption{Gate-dependent WAL-to-WL transitions in SnSe$_{2}$. (a) The $V_\text{g}$-dependent MR curves of SnSe$_{2}$ FET-SS3 at 2 K. (b) The corresponding MC characteristics fitted by the ILP theory. Note that the fitting is limited to $\pm1.5$ T, beyond which the conventional quadratic MR growth gradually takes over. (c) $H_{\varphi}$ and $H_\text{R2}$ extracted from the ILP theory fitting. (d) $\tau_\text{SO}$ extracted from the ILP theory fitting, showing $V_\text{g}$-polarity dependence of the inverse $\tau_\text{SO}$-$\tau_\text{p}$ relation.}
\label{fig4}
\end{figure}

It should be noticed that Eq. \ref{eqn:ILP} is a reduced form of the original ILP theory, with the prerequisite of predominant $\mathbf{k}$-cubic Rashba terms ($H_\text{R2}^{(3)}$) than the $\mathbf{k}$-linear Rashba SOC ($H_\text{R2}^{(1)}$) \cite{Cubic_Rashba_K_in_QWs, Cubic_Rashba_and_linear_K_in_asymmetric_Oxide_heterostructure}. In explicitly, $H_\text{R2}$ should be expressed as $H_\text{R2}=H_\text{R2}^{(1)}+H_\text{R2}^{(3)}$, in which $H_\text{R2}^{(3)}$ is sensitive to the Fermi energy ($E_\text{F}$) positions due to its cubic dependence on $\mathbf{k}$ \cite{gate_tunned_InSe}. The leading contribution of cubic Rashba SOC to the WAL-to-WL transitions is verified by gate-dependent MR study, which dynamically reveals the evolutions of WAL-to-WL crossovers as a function of $E_\text{F}$. Figure \ref{fig4}(a) summarizes the gate voltage ($V_\text{g}$)-dependent transport results of SnSe$_{2}$ FET-SS3 within the range of $\pm50$ V. For positive $V_\text{g}$, which enlarges the Fermi surface areas by introducing more electron doping, the ILP theory analyses indeed reveal a linear decrease in $\tau_\text{SO}$, accompanied by the inverse growth of $\tau_\text{p}$. While for negative $V_\text{g}$, the inverse relation between $\tau_\text{SO}$ and $\tau_\text{p}$ exhibits a significant smaller slope. The changes are mainly due to a slower $\tau_\text{SO}$ growth for negative $V_\text{g}$, because the Hall signals are linear over the whole $V_\text{g}$ range \cite{Note1}. The divergent $V_\text{g}$-polarity dependence of $\tau_\text{SO}$ can be understood by the screening of the positive-$V_\text{g}$ electric field due to the accumulation of additional holes at the interfacial regime, while the negative-$V_\text{g}$ electric field is able to penetrate more SnSe$_{2}$ layers by the depletion mechanism \cite{lujiong_BP_nanoletter}. By introducing a coefficient $r=H_\text{R2}^{(1)}/H_\text{R2}^{(3)}$ \cite{Te_WAL_PRB}, we have also fitted the $V_\text{g}$-dependent WAL-to-WL crossovers in Figure \ref{fig4}(b) by the original ILP theory, which further confirms the dominant role of $H_\text{R2}^{(3)}$ with r $<$ 1/3. Such a finding also explains the DFT calculated hidden spin textures in Figure \ref{fig1}(c), which are conspicuously deviating the helical spin-momentum locking structure of linear $\mathbf{k}$-Rashba. 

In summary, we have demonstrated that by introducing Se vacancies, the resulting SOC scattering can effectively create spin polarizations with finite spin relaxation time in multilayer 1T-SnSe$_{2}$. The surprising finding is rooted in the motion-narrowing random walks of processing conduction electrons in the strong build-in R2-Rashba fields, when the chiral symmetry between the two Se-sublattices is broken by vacancy-site SOC scattering which induces out-of-plane magnetizations for conduction electrons. Although the pronounced WAL-to-WL crossover behaviors in our quantum transport data can be excellently modeled by the ILP theory with a microscopic DP spin relaxation mechanism, the leading contributions of cubic Rashba SOC suggest that our current understanding on the hidden spin mechanisms may be worthy of further explorations. 


\begin{acknowledgments}
This work is supported by the Zhejiang Provincial Natural Science Foundation (Grant No. D19A040001), the National Science Foundation of China (Grant No. 11790313), and the National Key R\&D Program of the MOST of China (Grant Nos. 2017YFA0303002). A portion of this work was performed on the Steady High Magnetic Field Facilities, High Magnetic Field Laboratory, CAS.
\end{acknowledgments}


%


\begin{thebibliography}{36}%
\makeatletter
\providecommand \@ifxundefined [1]{%
 \@ifx{#1\undefined}
}%
\providecommand \@ifnum [1]{%
 \ifnum #1\expandafter \@firstoftwo
 \else \expandafter \@secondoftwo
 \fi
}%
\providecommand \@ifx [1]{%
 \ifx #1\expandafter \@firstoftwo
 \else \expandafter \@secondoftwo
 \fi
}%
\providecommand \natexlab [1]{#1}%
\providecommand \enquote  [1]{``#1''}%
\providecommand \bibnamefont  [1]{#1}%
\providecommand \bibfnamefont [1]{#1}%
\providecommand \citenamefont [1]{#1}%
\providecommand \href@noop [0]{\@secondoftwo}%
\providecommand \href [0]{\begingroup \@sanitize@url \@href}%
\providecommand \@href[1]{\@@startlink{#1}\@@href}%
\providecommand \@@href[1]{\endgroup#1\@@endlink}%
\providecommand \@sanitize@url [0]{\catcode `\\12\catcode `\$12\catcode
  `\&12\catcode `\#12\catcode `\^12\catcode `\_12\catcode `\%12\relax}%
\providecommand \@@startlink[1]{}%
\providecommand \@@endlink[0]{}%
\providecommand \url  [0]{\begingroup\@sanitize@url \@url }%
\providecommand \@url [1]{\endgroup\@href {#1}{\urlprefix }}%
\providecommand \urlprefix  [0]{URL }%
\providecommand \Eprint [0]{\href }%
\providecommand \doibase [0]{http://dx.doi.org/}%
\providecommand \selectlanguage [0]{\@gobble}%
\providecommand \bibinfo  [0]{\@secondoftwo}%
\providecommand \bibfield  [0]{\@secondoftwo}%
\providecommand \translation [1]{[#1]}%
\providecommand \BibitemOpen [0]{}%
\providecommand \bibitemStop [0]{}%
\providecommand \bibitemNoStop [0]{.\EOS\space}%
\providecommand \EOS [0]{\spacefactor3000\relax}%
\providecommand \BibitemShut  [1]{\csname bibitem#1\endcsname}%
\let\auto@bib@innerbib\@empty
\bibitem [{\citenamefont {Winkler}(2003)}]{winkler_SOC_book}%
  \BibitemOpen
  \bibfield  {author} {\bibinfo {author} {\bibfnamefont {R.}~\bibnamefont
  {Winkler}},\ }in\ \href {\doibase https://doi.org/10.1007/b13586} {\emph
  {\bibinfo {booktitle} {Spin-orbit Coupling Effects in Two-Dimensional
  Electron and Hole Systems}}}\ (\bibinfo  {publisher} {Springer Berlin,
  Heidelberg},\ \bibinfo {year} {2003})\BibitemShut {NoStop}%
\bibitem [{\citenamefont {Manchon}\ \emph {et~al.}(2015)\citenamefont
  {Manchon}, \citenamefont {Koo}, \citenamefont {Nitta}, \citenamefont
  {Frolov},\ and\ \citenamefont {Duine}}]{Rashba_SOC_review}%
  \BibitemOpen
  \bibfield  {author} {\bibinfo {author} {\bibfnamefont {A.}~\bibnamefont
  {Manchon}}, \bibinfo {author} {\bibfnamefont {H.~C.}\ \bibnamefont {Koo}},
  \bibinfo {author} {\bibfnamefont {J.}~\bibnamefont {Nitta}}, \bibinfo
  {author} {\bibfnamefont {S.~M.}\ \bibnamefont {Frolov}}, \ and\ \bibinfo
  {author} {\bibfnamefont {R.~A.}\ \bibnamefont {Duine}},\ }\href {\doibase
  https://doi.org/10.1038/nmat4360} {\bibfield  {journal} {\bibinfo  {journal}
  {Nat.\ Mater.}\ }\textbf {\bibinfo {volume} {14}},\ \bibinfo {pages} {871}
  (\bibinfo {year} {2015})}\BibitemShut {NoStop}%
\bibitem [{\citenamefont {Dresselhaus}(1955)}]{Dresselhaus_PR_1955}%
  \BibitemOpen
  \bibfield  {author} {\bibinfo {author} {\bibfnamefont {G.~F.}\ \bibnamefont
  {Dresselhaus}},\ }\href {\doibase https://doi.org/10.1103/PhysRev.100.580}
  {\bibfield  {journal} {\bibinfo  {journal} {Phys. Rev.}\ }\textbf {\bibinfo
  {volume} {100}},\ \bibinfo {pages} {580} (\bibinfo {year}
  {1955})}\BibitemShut {NoStop}%
\bibitem [{\citenamefont {Rashba}(1960)}]{rashba_1960_properties}%
  \BibitemOpen
  \bibfield  {author} {\bibinfo {author} {\bibfnamefont {E.~I.}\ \bibnamefont
  {Rashba}},\ }\href@noop {} {\bibfield  {journal} {\bibinfo  {journal} {Sov.
  Phys. Solid State}\ }\textbf {\bibinfo {volume} {2}},\ \bibinfo {pages}
  {1224} (\bibinfo {year} {1960})}\BibitemShut {NoStop}%
\bibitem [{\citenamefont {Zhang}\ \emph {et~al.}(2014)\citenamefont {Zhang},
  \citenamefont {Liu}, \citenamefont {Luo}, \citenamefont {Freeman},\ and\
  \citenamefont {Zunger}}]{hidden_spin_origin_calculation}%
  \BibitemOpen
  \bibfield  {author} {\bibinfo {author} {\bibfnamefont {X.}~\bibnamefont
  {Zhang}}, \bibinfo {author} {\bibfnamefont {Q.}~\bibnamefont {Liu}}, \bibinfo
  {author} {\bibfnamefont {J.}~\bibnamefont {Luo}}, \bibinfo {author}
  {\bibfnamefont {A.~J.}\ \bibnamefont {Freeman}}, \ and\ \bibinfo {author}
  {\bibfnamefont {A.}~\bibnamefont {Zunger}},\ }\href {\doibase
  https://doi.org/10.1038/nphys2933} {\bibfield  {journal} {\bibinfo  {journal}
  {Nat.\ Phys.}\ }\textbf {\bibinfo {volume} {10}},\ \bibinfo {pages} {387}
  (\bibinfo {year} {2014})}\BibitemShut {NoStop}%
\bibitem [{\citenamefont {Riley}\ \emph {et~al.}(2014)\citenamefont {Riley},
  \citenamefont {Mazzola}, \citenamefont {Dendzik}, \citenamefont {Michiardi},
  \citenamefont {Takayama}, \citenamefont {Bawden}, \citenamefont
  {Graner{\o}d}, \citenamefont {Leandersson}, \citenamefont {Balasubramanian},
  \citenamefont {Hoesch}, \citenamefont {Kim}, \citenamefont {Takagi},
  \citenamefont {Meevasana}, \citenamefont {Hofmann}, \citenamefont {Bahramy},
  \citenamefont {Wells},\ and\ \citenamefont {King}}]{bulk_WSe2_hidden_spin}%
  \BibitemOpen
  \bibfield  {author} {\bibinfo {author} {\bibfnamefont {J.~M.}\ \bibnamefont
  {Riley}}, \bibinfo {author} {\bibfnamefont {F.}~\bibnamefont {Mazzola}},
  \bibinfo {author} {\bibfnamefont {M.}~\bibnamefont {Dendzik}}, \bibinfo
  {author} {\bibfnamefont {M.}~\bibnamefont {Michiardi}}, \bibinfo {author}
  {\bibfnamefont {T.}~\bibnamefont {Takayama}}, \bibinfo {author}
  {\bibfnamefont {L.}~\bibnamefont {Bawden}}, \bibinfo {author} {\bibfnamefont
  {C.}~\bibnamefont {Graner{\o}d}}, \bibinfo {author} {\bibfnamefont
  {M.}~\bibnamefont {Leandersson}}, \bibinfo {author} {\bibfnamefont
  {T.}~\bibnamefont {Balasubramanian}}, \bibinfo {author} {\bibfnamefont
  {M.}~\bibnamefont {Hoesch}}, \bibinfo {author} {\bibfnamefont {T.~K.}\
  \bibnamefont {Kim}}, \bibinfo {author} {\bibfnamefont {H.}~\bibnamefont
  {Takagi}}, \bibinfo {author} {\bibfnamefont {W.}~\bibnamefont {Meevasana}},
  \bibinfo {author} {\bibfnamefont {P.}~\bibnamefont {Hofmann}}, \bibinfo
  {author} {\bibfnamefont {M.~S.}\ \bibnamefont {Bahramy}}, \bibinfo {author}
  {\bibfnamefont {J.~W.}\ \bibnamefont {Wells}}, \ and\ \bibinfo {author}
  {\bibfnamefont {P.~D.~C.}\ \bibnamefont {King}},\ }\href {\doibase
  https://doi.org/10.1038/NPHYS3105} {\bibfield  {journal} {\bibinfo  {journal}
  {Nat.\ Phys.}\ }\textbf {\bibinfo {volume} {10}},\ \bibinfo {pages} {835}
  (\bibinfo {year} {2014})}\BibitemShut {NoStop}%
\bibitem [{\citenamefont {Yao}\ \emph {et~al.}(2017)\citenamefont {Yao},
  \citenamefont {Wang}, \citenamefont {Huang}, \citenamefont {Deng},
  \citenamefont {Yan}, \citenamefont {Zhang}, \citenamefont {Miyamoto},
  \citenamefont {Okuda}, \citenamefont {Li}, \citenamefont {Wang},
  \citenamefont {Gao}, \citenamefont {Liu}, \citenamefont {Duan},\ and\
  \citenamefont {Zhou}}]{PtSe2_hidden_spin_Arpes}%
  \BibitemOpen
  \bibfield  {author} {\bibinfo {author} {\bibfnamefont {W.}~\bibnamefont
  {Yao}}, \bibinfo {author} {\bibfnamefont {E.}~\bibnamefont {Wang}}, \bibinfo
  {author} {\bibfnamefont {H.}~\bibnamefont {Huang}}, \bibinfo {author}
  {\bibfnamefont {K.}~\bibnamefont {Deng}}, \bibinfo {author} {\bibfnamefont
  {M.}~\bibnamefont {Yan}}, \bibinfo {author} {\bibfnamefont {K.}~\bibnamefont
  {Zhang}}, \bibinfo {author} {\bibfnamefont {K.}~\bibnamefont {Miyamoto}},
  \bibinfo {author} {\bibfnamefont {T.}~\bibnamefont {Okuda}}, \bibinfo
  {author} {\bibfnamefont {L.}~\bibnamefont {Li}}, \bibinfo {author}
  {\bibfnamefont {Y.}~\bibnamefont {Wang}}, \bibinfo {author} {\bibfnamefont
  {H.}~\bibnamefont {Gao}}, \bibinfo {author} {\bibfnamefont {C.}~\bibnamefont
  {Liu}}, \bibinfo {author} {\bibfnamefont {W.}~\bibnamefont {Duan}}, \ and\
  \bibinfo {author} {\bibfnamefont {S.}~\bibnamefont {Zhou}},\ }\href {\doibase
  https://doi.org/10.1038/ncomms14216} {\bibfield  {journal} {\bibinfo
  {journal} {Nat.\ Commun.}\ }\textbf {\bibinfo {volume} {8}},\ \bibinfo
  {pages} {1} (\bibinfo {year} {2017})}\BibitemShut {NoStop}%
\bibitem [{\citenamefont {Shafique}\ \emph {et~al.}(2017)\citenamefont
  {Shafique}, \citenamefont {Samad},\ and\ \citenamefont
  {Shin}}]{thermoelectric}%
  \BibitemOpen
  \bibfield  {author} {\bibinfo {author} {\bibfnamefont {A.}~\bibnamefont
  {Shafique}}, \bibinfo {author} {\bibfnamefont {A.}~\bibnamefont {Samad}}, \
  and\ \bibinfo {author} {\bibfnamefont {Y.}~\bibnamefont {Shin}},\ }\href
  {\doibase https://doi.org/10.1039/c7cp03748a} {\bibfield  {journal} {\bibinfo
   {journal} {Phys. Chem. Chem. Phys}\ }\textbf {\bibinfo {volume} {19}},\
  \bibinfo {pages} {20677} (\bibinfo {year} {2017})}\BibitemShut {NoStop}%
\bibitem [{\citenamefont {Zhou}\ \emph {et~al.}(2017)\citenamefont {Zhou},
  \citenamefont {Zhou}, \citenamefont {Li}, \citenamefont {Song}, \citenamefont
  {Zhang}, \citenamefont {Hu}, \citenamefont {Gan}, \citenamefont {Li},
  \citenamefont {L{\"u}}, \citenamefont {Luo}, \citenamefont {Xiong},\ and\
  \citenamefont {Zhai}}]{photodetectors}%
  \BibitemOpen
  \bibfield  {author} {\bibinfo {author} {\bibfnamefont {X.}~\bibnamefont
  {Zhou}}, \bibinfo {author} {\bibfnamefont {N.}~\bibnamefont {Zhou}}, \bibinfo
  {author} {\bibfnamefont {C.}~\bibnamefont {Li}}, \bibinfo {author}
  {\bibfnamefont {H.}~\bibnamefont {Song}}, \bibinfo {author} {\bibfnamefont
  {Q.}~\bibnamefont {Zhang}}, \bibinfo {author} {\bibfnamefont
  {X.}~\bibnamefont {Hu}}, \bibinfo {author} {\bibfnamefont {L.}~\bibnamefont
  {Gan}}, \bibinfo {author} {\bibfnamefont {H.}~\bibnamefont {Li}}, \bibinfo
  {author} {\bibfnamefont {J.}~\bibnamefont {L{\"u}}}, \bibinfo {author}
  {\bibfnamefont {J.}~\bibnamefont {Luo}}, \bibinfo {author} {\bibfnamefont
  {J.}~\bibnamefont {Xiong}}, \ and\ \bibinfo {author} {\bibfnamefont
  {T.}~\bibnamefont {Zhai}},\ }\href {\doibase
  https://doi.org/10.1088/2053-1583/aa6422} {\bibfield  {journal} {\bibinfo
  {journal} {2D Mater.}\ }\textbf {\bibinfo {volume} {4}},\ \bibinfo {pages}
  {025048} (\bibinfo {year} {2017})}\BibitemShut {NoStop}%
\bibitem [{\citenamefont {Tan}\ \emph {et~al.}(2017)\citenamefont {Tan},
  \citenamefont {Chen}, \citenamefont {Wu}, \citenamefont {Shang},
  \citenamefont {Liu}, \citenamefont {Pan},\ and\ \citenamefont
  {Xiong}}]{photocatalyst}%
  \BibitemOpen
  \bibfield  {author} {\bibinfo {author} {\bibfnamefont {P.}~\bibnamefont
  {Tan}}, \bibinfo {author} {\bibfnamefont {X.}~\bibnamefont {Chen}}, \bibinfo
  {author} {\bibfnamefont {L.}~\bibnamefont {Wu}}, \bibinfo {author}
  {\bibfnamefont {Y.}~\bibnamefont {Shang}}, \bibinfo {author} {\bibfnamefont
  {W.}~\bibnamefont {Liu}}, \bibinfo {author} {\bibfnamefont {J.}~\bibnamefont
  {Pan}}, \ and\ \bibinfo {author} {\bibfnamefont {X.}~\bibnamefont {Xiong}},\
  }\href {\doibase https://doi.org/10.1016/j.apcatb.2016.09.033} {\bibfield
  {journal} {\bibinfo  {journal} {Appl. Catal. B: Environ.}\ }\textbf {\bibinfo
  {volume} {202}},\ \bibinfo {pages} {326} (\bibinfo {year}
  {2017})}\BibitemShut {NoStop}%
\bibitem [{\citenamefont {Datta}\ and\ \citenamefont
  {Das}(1990)}]{FET_spin_transport}%
  \BibitemOpen
  \bibfield  {author} {\bibinfo {author} {\bibfnamefont {S.}~\bibnamefont
  {Datta}}\ and\ \bibinfo {author} {\bibfnamefont {B.}~\bibnamefont {Das}},\
  }\href {\doibase https://doi.org/10.1063/1.102730} {\bibfield  {journal}
  {\bibinfo  {journal} {Appl.\ Phys.\ Lett.}\ }\textbf {\bibinfo {volume}
  {56}},\ \bibinfo {pages} {665} (\bibinfo {year} {1990})}\BibitemShut
  {NoStop}%
\bibitem [{\citenamefont {Zutic}\ \emph {et~al.}(2004)\citenamefont {Zutic},
  \citenamefont {Fabian},\ and\ \citenamefont
  {Das~Sarma}}]{Origin_spintronics_review}%
  \BibitemOpen
  \bibfield  {author} {\bibinfo {author} {\bibfnamefont {I.}~\bibnamefont
  {Zutic}}, \bibinfo {author} {\bibfnamefont {J.}~\bibnamefont {Fabian}}, \
  and\ \bibinfo {author} {\bibfnamefont {S.}~\bibnamefont {Das~Sarma}},\ }\href
  {\doibase https://doi.org/10.1103/revmodphys.76.323} {\bibfield  {journal}
  {\bibinfo  {journal} {Rev. Mod. Phys.}\ }\textbf {\bibinfo {volume} {76}},\
  \bibinfo {pages} {323} (\bibinfo {year} {2004})}\BibitemShut {NoStop}%
\bibitem [{\citenamefont {Han}\ \emph {et~al.}(2014)\citenamefont {Han},
  \citenamefont {Kawakami}, \citenamefont {Gmitra},\ and\ \citenamefont
  {Fabian}}]{graphene_spintronics}%
  \BibitemOpen
  \bibfield  {author} {\bibinfo {author} {\bibfnamefont {W.}~\bibnamefont
  {Han}}, \bibinfo {author} {\bibfnamefont {R.~K.}\ \bibnamefont {Kawakami}},
  \bibinfo {author} {\bibfnamefont {M.}~\bibnamefont {Gmitra}}, \ and\ \bibinfo
  {author} {\bibfnamefont {J.}~\bibnamefont {Fabian}},\ }\href {\doibase
  https://doi.org/10.1038/nnano.2014.214} {\bibfield  {journal} {\bibinfo
  {journal} {Nat.\ Nanotech.}\ }\textbf {\bibinfo {volume} {9}},\ \bibinfo
  {pages} {794} (\bibinfo {year} {2014})}\BibitemShut {NoStop}%
\bibitem [{\citenamefont {Schmidt}\ \emph {et~al.}(2016)\citenamefont
  {Schmidt}, \citenamefont {Yudhistira}, \citenamefont {Chu}, \citenamefont
  {Castro~Neto}, \citenamefont {\"Ozyilmaz}, \citenamefont {Adam},\ and\
  \citenamefont {Eda}}]{DP_MoS2_PRL}%
  \BibitemOpen
  \bibfield  {author} {\bibinfo {author} {\bibfnamefont {H.}~\bibnamefont
  {Schmidt}}, \bibinfo {author} {\bibfnamefont {I.}~\bibnamefont {Yudhistira}},
  \bibinfo {author} {\bibfnamefont {L.}~\bibnamefont {Chu}}, \bibinfo {author}
  {\bibfnamefont {A.~H.}\ \bibnamefont {Castro~Neto}}, \bibinfo {author}
  {\bibfnamefont {B.}~\bibnamefont {\"Ozyilmaz}}, \bibinfo {author}
  {\bibfnamefont {S.}~\bibnamefont {Adam}}, \ and\ \bibinfo {author}
  {\bibfnamefont {G.}~\bibnamefont {Eda}},\ }\href {\doibase
  https://doi.org/10.1103/PhysRevLett.116.046803} {\bibfield  {journal}
  {\bibinfo  {journal} {Phys.\ Rev.\ Lett.}\ }\textbf {\bibinfo {volume}
  {116}},\ \bibinfo {pages} {046803} (\bibinfo {year} {2016})}\BibitemShut
  {NoStop}%
\bibitem [{\citenamefont {Wang}\ \emph {et~al.}(2018)\citenamefont {Wang},
  \citenamefont {Fan}, \citenamefont {Shen}, \citenamefont {Hua}, \citenamefont
  {Hu}, \citenamefont {Sheng}, \citenamefont {Lu}, \citenamefont {Fang},
  \citenamefont {Qiu}, \citenamefont {Lu}, \citenamefont {Liu}, \citenamefont
  {Liu}, \citenamefont {Huang}, \citenamefont {Xu}, \citenamefont {Shen},\ and\
  \citenamefont {Zheng}}]{wangzhen_SnSe_NC}%
  \BibitemOpen
  \bibfield  {author} {\bibinfo {author} {\bibfnamefont {Z.}~\bibnamefont
  {Wang}}, \bibinfo {author} {\bibfnamefont {C.}~\bibnamefont {Fan}}, \bibinfo
  {author} {\bibfnamefont {Z.}~\bibnamefont {Shen}}, \bibinfo {author}
  {\bibfnamefont {C.}~\bibnamefont {Hua}}, \bibinfo {author} {\bibfnamefont
  {Q.}~\bibnamefont {Hu}}, \bibinfo {author} {\bibfnamefont {F.}~\bibnamefont
  {Sheng}}, \bibinfo {author} {\bibfnamefont {Y.}~\bibnamefont {Lu}}, \bibinfo
  {author} {\bibfnamefont {H.}~\bibnamefont {Fang}}, \bibinfo {author}
  {\bibfnamefont {Z.}~\bibnamefont {Qiu}}, \bibinfo {author} {\bibfnamefont
  {J.}~\bibnamefont {Lu}}, \bibinfo {author} {\bibfnamefont {Z.}~\bibnamefont
  {Liu}}, \bibinfo {author} {\bibfnamefont {W.}~\bibnamefont {Liu}}, \bibinfo
  {author} {\bibfnamefont {Y.}~\bibnamefont {Huang}}, \bibinfo {author}
  {\bibfnamefont {Z.}~\bibnamefont {Xu}}, \bibinfo {author} {\bibfnamefont
  {D.~W.}\ \bibnamefont {Shen}}, \ and\ \bibinfo {author} {\bibfnamefont
  {Y.}~\bibnamefont {Zheng}},\ }\href {\doibase
  https://doi.org/10.1038/s41467-017-02566-1} {\bibfield  {journal} {\bibinfo
  {journal} {Nat.\ Commun.}\ }\textbf {\bibinfo {volume} {9}},\ \bibinfo
  {pages} {1} (\bibinfo {year} {2018})}\BibitemShut {NoStop}%
\bibitem [{Note1()}]{Note1}%
  \BibitemOpen
  \bibinfo {note} {See Supplemental Material at
  http://link.aps.org/supplemetal/ for Raman and AFM characterizations, DFT
  calculations of R2-Rashba hidden spins and electron doping by Se
  vacancies.}\BibitemShut {Stop}%
\bibitem [{\citenamefont
  {Dyakonov}(2008)}]{Dyakonov_Spin_Physics_in_Semiconductors}%
  \BibitemOpen
  \bibfield  {author} {\bibinfo {author} {\bibfnamefont {M.~I.}\ \bibnamefont
  {Dyakonov}},\ }in\ \href {\doibase https://doi.org/10.1007/978-3-540-78820-1}
  {\emph {\bibinfo {booktitle} {Spin Physics in Semiconductors}}}\ (\bibinfo
  {publisher} {Springer Berlin, Heidelberg},\ \bibinfo {year}
  {2008})\BibitemShut {NoStop}%
\bibitem [{\citenamefont {Balakrishnan}\ \emph {et~al.}(2013)\citenamefont
  {Balakrishnan}, \citenamefont {Koon}, \citenamefont {Jaiswal}, \citenamefont
  {Castro~Neto},\ and\ \citenamefont {{\"O}zyilmaz}}]{H_graphene_SOC_Barbaros}%
  \BibitemOpen
  \bibfield  {author} {\bibinfo {author} {\bibfnamefont {J.}~\bibnamefont
  {Balakrishnan}}, \bibinfo {author} {\bibfnamefont {G.~K.~W.}\ \bibnamefont
  {Koon}}, \bibinfo {author} {\bibfnamefont {M.}~\bibnamefont {Jaiswal}},
  \bibinfo {author} {\bibfnamefont {A.~H.}\ \bibnamefont {Castro~Neto}}, \ and\
  \bibinfo {author} {\bibfnamefont {B.}~\bibnamefont {{\"O}zyilmaz}},\ }\href
  {\doibase https://doi.org/10.1038/nphys2576} {\bibfield  {journal} {\bibinfo
  {journal} {Nat.\ Phys.}\ }\textbf {\bibinfo {volume} {9}},\ \bibinfo {pages}
  {284} (\bibinfo {year} {2013})}\BibitemShut {NoStop}%
\bibitem [{\citenamefont {Liu}\ \emph {et~al.}(2012)\citenamefont {Liu},
  \citenamefont {Zhang}, \citenamefont {Chang}, \citenamefont {Zhang},
  \citenamefont {Feng}, \citenamefont {Li}, \citenamefont {He}, \citenamefont
  {Wang}, \citenamefont {Chen}, \citenamefont {Dai}, \citenamefont {Fang},
  \citenamefont {Xue}, \citenamefont {Ma},\ and\ \citenamefont
  {Wang}}]{wangyayu_PRL_WALWL_crossover}%
  \BibitemOpen
  \bibfield  {author} {\bibinfo {author} {\bibfnamefont {M.}~\bibnamefont
  {Liu}}, \bibinfo {author} {\bibfnamefont {J.}~\bibnamefont {Zhang}}, \bibinfo
  {author} {\bibfnamefont {C.}~\bibnamefont {Chang}}, \bibinfo {author}
  {\bibfnamefont {Z.}~\bibnamefont {Zhang}}, \bibinfo {author} {\bibfnamefont
  {X.}~\bibnamefont {Feng}}, \bibinfo {author} {\bibfnamefont {K.}~\bibnamefont
  {Li}}, \bibinfo {author} {\bibfnamefont {K.}~\bibnamefont {He}}, \bibinfo
  {author} {\bibfnamefont {L.}~\bibnamefont {Wang}}, \bibinfo {author}
  {\bibfnamefont {X.}~\bibnamefont {Chen}}, \bibinfo {author} {\bibfnamefont
  {X.}~\bibnamefont {Dai}}, \bibinfo {author} {\bibfnamefont {Z.}~\bibnamefont
  {Fang}}, \bibinfo {author} {\bibfnamefont {Q.}~\bibnamefont {Xue}}, \bibinfo
  {author} {\bibfnamefont {X.}~\bibnamefont {Ma}}, \ and\ \bibinfo {author}
  {\bibfnamefont {Y.}~\bibnamefont {Wang}},\ }\href {\doibase
  https://doi.org/10.1103/PhysRevLett.108.036805} {\bibfield  {journal}
  {\bibinfo  {journal} {Phys.\ Rev.\ Lett.}\ }\textbf {\bibinfo {volume}
  {108}},\ \bibinfo {pages} {036805} (\bibinfo {year} {2012})}\BibitemShut
  {NoStop}%
\bibitem [{\citenamefont {Wang}\ \emph {et~al.}(2016)\citenamefont {Wang},
  \citenamefont {Zheng}, \citenamefont {Shen}, \citenamefont {Lu},
  \citenamefont {Fang}, \citenamefont {Sheng}, \citenamefont {Zhou},
  \citenamefont {Yang}, \citenamefont {Li}, \citenamefont {Feng},\ and\
  \citenamefont {Xu}}]{WZ_PRB_WAL}%
  \BibitemOpen
  \bibfield  {author} {\bibinfo {author} {\bibfnamefont {Z.}~\bibnamefont
  {Wang}}, \bibinfo {author} {\bibfnamefont {Y.}~\bibnamefont {Zheng}},
  \bibinfo {author} {\bibfnamefont {Z.}~\bibnamefont {Shen}}, \bibinfo {author}
  {\bibfnamefont {Y.~H.}\ \bibnamefont {Lu}}, \bibinfo {author} {\bibfnamefont
  {H.}~\bibnamefont {Fang}}, \bibinfo {author} {\bibfnamefont {F.}~\bibnamefont
  {Sheng}}, \bibinfo {author} {\bibfnamefont {Y.}~\bibnamefont {Zhou}},
  \bibinfo {author} {\bibfnamefont {X.}~\bibnamefont {Yang}}, \bibinfo {author}
  {\bibfnamefont {Y.}~\bibnamefont {Li}}, \bibinfo {author} {\bibfnamefont
  {C.}~\bibnamefont {Feng}}, \ and\ \bibinfo {author} {\bibfnamefont {Z.~A.}\
  \bibnamefont {Xu}},\ }\href {\doibase
  https://doi.org/10.1103/PhysRevB.93.121112} {\bibfield  {journal} {\bibinfo
  {journal} {Phys.\ Rev.\ B}\ }\textbf {\bibinfo {volume} {93}},\ \bibinfo
  {pages} {121112(R)} (\bibinfo {year} {2016})}\BibitemShut {NoStop}%
\bibitem [{\citenamefont {Chen}\ \emph {et~al.}(2010)\citenamefont {Chen},
  \citenamefont {Qin}, \citenamefont {Yang}, \citenamefont {Liu}, \citenamefont
  {Guan}, \citenamefont {Qu}, \citenamefont {Zhang}, \citenamefont {Shi},
  \citenamefont {Xie}, \citenamefont {Yang}, \citenamefont {Wu}, \citenamefont
  {Li},\ and\ \citenamefont {Lu}}]{WAL_luli_prl}%
  \BibitemOpen
  \bibfield  {author} {\bibinfo {author} {\bibfnamefont {J.}~\bibnamefont
  {Chen}}, \bibinfo {author} {\bibfnamefont {H.~J.}\ \bibnamefont {Qin}},
  \bibinfo {author} {\bibfnamefont {F.}~\bibnamefont {Yang}}, \bibinfo {author}
  {\bibfnamefont {J.}~\bibnamefont {Liu}}, \bibinfo {author} {\bibfnamefont
  {T.}~\bibnamefont {Guan}}, \bibinfo {author} {\bibfnamefont {F.~M.}\
  \bibnamefont {Qu}}, \bibinfo {author} {\bibfnamefont {G.~H.}\ \bibnamefont
  {Zhang}}, \bibinfo {author} {\bibfnamefont {J.~R.}\ \bibnamefont {Shi}},
  \bibinfo {author} {\bibfnamefont {X.~C.}\ \bibnamefont {Xie}}, \bibinfo
  {author} {\bibfnamefont {C.~L.}\ \bibnamefont {Yang}}, \bibinfo {author}
  {\bibfnamefont {K.~H.}\ \bibnamefont {Wu}}, \bibinfo {author} {\bibfnamefont
  {Y.~Q.}\ \bibnamefont {Li}}, \ and\ \bibinfo {author} {\bibfnamefont
  {L.}~\bibnamefont {Lu}},\ }\href {\doibase
  https://doi.org/10.1103/PhysRevLett.105.176602} {\bibfield  {journal}
  {\bibinfo  {journal} {Phys.\ Rev.\ Lett.}\ }\textbf {\bibinfo {volume}
  {105}},\ \bibinfo {pages} {176602} (\bibinfo {year} {2010})}\BibitemShut
  {NoStop}%
\bibitem [{\citenamefont {Lu}\ and\ \citenamefont
  {Shen}(2011)}]{WAL_calculation_luahizhou_1}%
  \BibitemOpen
  \bibfield  {author} {\bibinfo {author} {\bibfnamefont {H.}~\bibnamefont
  {Lu}}\ and\ \bibinfo {author} {\bibfnamefont {S.}~\bibnamefont {Shen}},\
  }\href {\doibase https://doi.org/10.1103/PhysRevB.84.125138} {\bibfield
  {journal} {\bibinfo  {journal} {Phys.\ Rev.\ B}\ }\textbf {\bibinfo {volume}
  {84}},\ \bibinfo {pages} {125138} (\bibinfo {year} {2011})}\BibitemShut
  {NoStop}%
\bibitem [{\citenamefont {Lu}\ \emph {et~al.}(2011)\citenamefont {Lu},
  \citenamefont {Shi},\ and\ \citenamefont
  {Shen}}]{WAL_calculation_luahizhou_2}%
  \BibitemOpen
  \bibfield  {author} {\bibinfo {author} {\bibfnamefont {H.}~\bibnamefont
  {Lu}}, \bibinfo {author} {\bibfnamefont {J.}~\bibnamefont {Shi}}, \ and\
  \bibinfo {author} {\bibfnamefont {S.}~\bibnamefont {Shen}},\ }\href {\doibase
  https://doi.org/10.1103/PhysRevLett.107.076801} {\bibfield  {journal}
  {\bibinfo  {journal} {Phys.\ Rev.\ Lett.}\ }\textbf {\bibinfo {volume}
  {107}},\ \bibinfo {pages} {076801} (\bibinfo {year} {2011})}\BibitemShut
  {NoStop}%
\bibitem [{\citenamefont {Pikus}\ and\ \citenamefont {Pikus}(1995)}]{ILP_HLN}%
  \BibitemOpen
  \bibfield  {author} {\bibinfo {author} {\bibfnamefont {F.~G.}\ \bibnamefont
  {Pikus}}\ and\ \bibinfo {author} {\bibfnamefont {G.~E.}\ \bibnamefont
  {Pikus}},\ }\href {\doibase https://doi.org/10.1103/PhysRevB.51.16928}
  {\bibfield  {journal} {\bibinfo  {journal} {Phys.\ Rev.\ B}\ }\textbf
  {\bibinfo {volume} {51}},\ \bibinfo {pages} {16928} (\bibinfo {year}
  {1995})}\BibitemShut {NoStop}%
\bibitem [{Note2()}]{Note2}%
  \BibitemOpen
  \bibinfo {note} {$H_{\varphi }$ is correlated to the phase coherence time
  $\tau _{\varphi }$ by the relation $\tau _{\varphi } = m^{*}/4\pi \protect
  \hbar \mu nH_{\varphi }$. For quantum interference introduced conduction
  corrections, $H_{\varphi }$ is more intuitive than $\tau _{\varphi }$ for
  modeling and analyzing the experiment results, and the former is used for
  discussions in this letter.}\BibitemShut {Stop}%
\bibitem [{\citenamefont {D'yakonov}\ and\ \citenamefont
  {Perel}(1971)}]{DP_scattering}%
  \BibitemOpen
  \bibfield  {author} {\bibinfo {author} {\bibfnamefont {M.~I.}\ \bibnamefont
  {D'yakonov}}\ and\ \bibinfo {author} {\bibfnamefont {V.~I.}\ \bibnamefont
  {Perel}},\ }\href@noop {} {\bibfield  {journal} {\bibinfo  {journal} {Sov.
  Phys. JETP}\ }\textbf {\bibinfo {volume} {33}},\ \bibinfo {pages} {1053}
  (\bibinfo {year} {1971})}\BibitemShut {NoStop}%
\bibitem [{\citenamefont {Knap}\ \emph {et~al.}(1996)\citenamefont {Knap},
  \citenamefont {Skierbiszewski}, \citenamefont {Zduniak}, \citenamefont
  {Litwin-Staszewska}, \citenamefont {Bertho}, \citenamefont {Kobbi},
  \citenamefont {Robert}, \citenamefont {Pikus}, \citenamefont {Pikus},
  \citenamefont {Iordanskii}, \citenamefont {Mosser}, \citenamefont
  {Zekentes},\ and\ \citenamefont {Lyanda-Geller}}]{spin_procession_in_QWs}%
  \BibitemOpen
  \bibfield  {author} {\bibinfo {author} {\bibfnamefont {W.}~\bibnamefont
  {Knap}}, \bibinfo {author} {\bibfnamefont {C.}~\bibnamefont
  {Skierbiszewski}}, \bibinfo {author} {\bibfnamefont {A.}~\bibnamefont
  {Zduniak}}, \bibinfo {author} {\bibfnamefont {E.}~\bibnamefont
  {Litwin-Staszewska}}, \bibinfo {author} {\bibfnamefont {D.}~\bibnamefont
  {Bertho}}, \bibinfo {author} {\bibfnamefont {F.}~\bibnamefont {Kobbi}},
  \bibinfo {author} {\bibfnamefont {J.}~\bibnamefont {Robert}}, \bibinfo
  {author} {\bibfnamefont {G.}~\bibnamefont {Pikus}}, \bibinfo {author}
  {\bibfnamefont {F.}~\bibnamefont {Pikus}}, \bibinfo {author} {\bibfnamefont
  {S.}~\bibnamefont {Iordanskii}}, \bibinfo {author} {\bibfnamefont
  {V.}~\bibnamefont {Mosser}}, \bibinfo {author} {\bibfnamefont
  {K.}~\bibnamefont {Zekentes}}, \ and\ \bibinfo {author} {\bibfnamefont
  {Y.}~\bibnamefont {Lyanda-Geller}},\ }\href {\doibase
  https://doi.org/10.1103/PhysRevB.53.3912} {\bibfield  {journal} {\bibinfo
  {journal} {Phys.\ Rev.\ B}\ }\textbf {\bibinfo {volume} {53}},\ \bibinfo
  {pages} {3912} (\bibinfo {year} {1996})}\BibitemShut {NoStop}%
\bibitem [{\citenamefont {Elliott}(1954)}]{EY_E_scattering}%
  \BibitemOpen
  \bibfield  {author} {\bibinfo {author} {\bibfnamefont {R.~J.}\ \bibnamefont
  {Elliott}},\ }\href {\doibase https://doi.org/10.1103/PhysRev.96.266}
  {\bibfield  {journal} {\bibinfo  {journal} {Phys. Rev.}\ }\textbf {\bibinfo
  {volume} {96}},\ \bibinfo {pages} {266} (\bibinfo {year} {1954})}\BibitemShut
  {NoStop}%
\bibitem [{\citenamefont {Yafet}(1963)}]{EY_Y_scattering}%
  \BibitemOpen
  \bibfield  {author} {\bibinfo {author} {\bibfnamefont {Y.}~\bibnamefont
  {Yafet}},\ }in\ \href {\doibase
  https://doi.org/10.1016/S0081-1947(08)60259-3} {\emph {\bibinfo {booktitle}
  {g Factors and Spin-Lattice Relaxation of Conduction Electrons}}}\ (\bibinfo
  {publisher} {Academic Press},\ \bibinfo {year} {1963})\BibitemShut {NoStop}%
\bibitem [{\citenamefont {Hikami}\ \emph {et~al.}(1980)\citenamefont {Hikami},
  \citenamefont {Larkin},\ and\ \citenamefont {Nagaoka}}]{origin_HLN}%
  \BibitemOpen
  \bibfield  {author} {\bibinfo {author} {\bibfnamefont {S.}~\bibnamefont
  {Hikami}}, \bibinfo {author} {\bibfnamefont {A.~I.}\ \bibnamefont {Larkin}},
  \ and\ \bibinfo {author} {\bibfnamefont {Y.}~\bibnamefont {Nagaoka}},\ }\href
  {\doibase https://doi.org/10.1143/PTP.63.707} {\bibfield  {journal} {\bibinfo
   {journal} {Prog. Theor. Phys.}\ }\textbf {\bibinfo {volume} {63}},\ \bibinfo
  {pages} {707} (\bibinfo {year} {1980})}\BibitemShut {NoStop}%
\bibitem [{Note3()}]{Note3}%
  \BibitemOpen
  \bibinfo {note} {$m^{\ast }$ used here to calculate $\tau _\protect \text
  {p}$ is deduced from the density functional theory calculations, because the
  Shubnikov-de Haas oscillations are too weak to extract convincing
  informations.}\BibitemShut {Stop}%
\bibitem [{\citenamefont {Moriya}\ \emph {et~al.}(2014)\citenamefont {Moriya},
  \citenamefont {Sawano}, \citenamefont {Hoshi}, \citenamefont {Masubuchi},
  \citenamefont {Shiraki}, \citenamefont {Wild}, \citenamefont {Neumann},
  \citenamefont {Abstreiter}, \citenamefont {Bougeard}, \citenamefont {Koga},\
  and\ \citenamefont {Machida}}]{Cubic_Rashba_K_in_QWs}%
  \BibitemOpen
  \bibfield  {author} {\bibinfo {author} {\bibfnamefont {R.}~\bibnamefont
  {Moriya}}, \bibinfo {author} {\bibfnamefont {K.}~\bibnamefont {Sawano}},
  \bibinfo {author} {\bibfnamefont {Y.}~\bibnamefont {Hoshi}}, \bibinfo
  {author} {\bibfnamefont {S.}~\bibnamefont {Masubuchi}}, \bibinfo {author}
  {\bibfnamefont {Y.}~\bibnamefont {Shiraki}}, \bibinfo {author} {\bibfnamefont
  {A.}~\bibnamefont {Wild}}, \bibinfo {author} {\bibfnamefont {C.}~\bibnamefont
  {Neumann}}, \bibinfo {author} {\bibfnamefont {G.}~\bibnamefont {Abstreiter}},
  \bibinfo {author} {\bibfnamefont {D.}~\bibnamefont {Bougeard}}, \bibinfo
  {author} {\bibfnamefont {T.}~\bibnamefont {Koga}}, \ and\ \bibinfo {author}
  {\bibfnamefont {T.}~\bibnamefont {Machida}},\ }\href {\doibase
  https://doi.org/10.1103/PhysRevLett.113.086601} {\bibfield  {journal}
  {\bibinfo  {journal} {Phys.\ Rev.\ Lett.}\ }\textbf {\bibinfo {volume}
  {113}},\ \bibinfo {pages} {086601} (\bibinfo {year} {2014})}\BibitemShut
  {NoStop}%
\bibitem [{\citenamefont {Lin}\ \emph {et~al.}(2019)\citenamefont {Lin},
  \citenamefont {Li}, \citenamefont {Do{\u{g}}an}, \citenamefont {Li},
  \citenamefont {Rotella}, \citenamefont {Yu}, \citenamefont {Zhang},
  \citenamefont {Li}, \citenamefont {Lew}, \citenamefont {Wang}, \citenamefont
  {Prellier}, \citenamefont {Pennycook}, \citenamefont {Chen}, \citenamefont
  {Zhong}, \citenamefont {Manchon},\ and\ \citenamefont
  {Wu}}]{Cubic_Rashba_and_linear_K_in_asymmetric_Oxide_heterostructure}%
  \BibitemOpen
  \bibfield  {author} {\bibinfo {author} {\bibfnamefont {W.}~\bibnamefont
  {Lin}}, \bibinfo {author} {\bibfnamefont {L.}~\bibnamefont {Li}}, \bibinfo
  {author} {\bibfnamefont {F.}~\bibnamefont {Do{\u{g}}an}}, \bibinfo {author}
  {\bibfnamefont {C.}~\bibnamefont {Li}}, \bibinfo {author} {\bibfnamefont
  {H.}~\bibnamefont {Rotella}}, \bibinfo {author} {\bibfnamefont
  {X.}~\bibnamefont {Yu}}, \bibinfo {author} {\bibfnamefont {B.}~\bibnamefont
  {Zhang}}, \bibinfo {author} {\bibfnamefont {Y.}~\bibnamefont {Li}}, \bibinfo
  {author} {\bibfnamefont {W.~S.}\ \bibnamefont {Lew}}, \bibinfo {author}
  {\bibfnamefont {S.}~\bibnamefont {Wang}}, \bibinfo {author} {\bibfnamefont
  {W.}~\bibnamefont {Prellier}}, \bibinfo {author} {\bibfnamefont {S.~J.}\
  \bibnamefont {Pennycook}}, \bibinfo {author} {\bibfnamefont {J.}~\bibnamefont
  {Chen}}, \bibinfo {author} {\bibfnamefont {Z.}~\bibnamefont {Zhong}},
  \bibinfo {author} {\bibfnamefont {A.}~\bibnamefont {Manchon}}, \ and\
  \bibinfo {author} {\bibfnamefont {T.}~\bibnamefont {Wu}},\ }\href {\doibase
  https://doi.org/10.1038/s41467-019-10961-z} {\bibfield  {journal} {\bibinfo
  {journal} {Nat.\ Commun.}\ }\textbf {\bibinfo {volume} {10}},\ \bibinfo
  {pages} {1} (\bibinfo {year} {2019})}\BibitemShut {NoStop}%
\bibitem [{\citenamefont {Zeng}\ \emph {et~al.}(2018)\citenamefont {Zeng},
  \citenamefont {Liang}, \citenamefont {Gao}, \citenamefont {Wang},
  \citenamefont {Pan}, \citenamefont {Wu}, \citenamefont {Liu}, \citenamefont
  {Zhang}, \citenamefont {Cao}, \citenamefont {Liu}, \citenamefont {Fu},
  \citenamefont {Wang}, \citenamefont {Watanabe}, \citenamefont {Taniguchi},
  \citenamefont {Lu},\ and\ \citenamefont {Miao}}]{gate_tunned_InSe}%
  \BibitemOpen
  \bibfield  {author} {\bibinfo {author} {\bibfnamefont {J.}~\bibnamefont
  {Zeng}}, \bibinfo {author} {\bibfnamefont {S.-J.}\ \bibnamefont {Liang}},
  \bibinfo {author} {\bibfnamefont {A.}~\bibnamefont {Gao}}, \bibinfo {author}
  {\bibfnamefont {Y.}~\bibnamefont {Wang}}, \bibinfo {author} {\bibfnamefont
  {C.}~\bibnamefont {Pan}}, \bibinfo {author} {\bibfnamefont {C.}~\bibnamefont
  {Wu}}, \bibinfo {author} {\bibfnamefont {E.}~\bibnamefont {Liu}}, \bibinfo
  {author} {\bibfnamefont {L.}~\bibnamefont {Zhang}}, \bibinfo {author}
  {\bibfnamefont {T.}~\bibnamefont {Cao}}, \bibinfo {author} {\bibfnamefont
  {X.}~\bibnamefont {Liu}}, \bibinfo {author} {\bibfnamefont {Y.}~\bibnamefont
  {Fu}}, \bibinfo {author} {\bibfnamefont {Y.}~\bibnamefont {Wang}}, \bibinfo
  {author} {\bibfnamefont {K.}~\bibnamefont {Watanabe}}, \bibinfo {author}
  {\bibfnamefont {T.}~\bibnamefont {Taniguchi}}, \bibinfo {author}
  {\bibfnamefont {H.}~\bibnamefont {Lu}}, \ and\ \bibinfo {author}
  {\bibfnamefont {F.}~\bibnamefont {Miao}},\ }\href {\doibase
  https://doi.org/10.1103/PhysRevB.98.125414} {\bibfield  {journal} {\bibinfo
  {journal} {Phys.\ Rev.\ B}\ }\textbf {\bibinfo {volume} {98}},\ \bibinfo
  {pages} {125414} (\bibinfo {year} {2018})}\BibitemShut {NoStop}%
\bibitem [{\citenamefont {Liu}\ \emph {et~al.}(2017)\citenamefont {Liu},
  \citenamefont {Qiu}, \citenamefont {Carvalho}, \citenamefont {Bao},
  \citenamefont {Xu}, \citenamefont {Tan}, \citenamefont {Liu}, \citenamefont
  {Castro~Neto}, \citenamefont {Loh},\ and\ \citenamefont
  {Lu}}]{lujiong_BP_nanoletter}%
  \BibitemOpen
  \bibfield  {author} {\bibinfo {author} {\bibfnamefont {Y.}~\bibnamefont
  {Liu}}, \bibinfo {author} {\bibfnamefont {Z.}~\bibnamefont {Qiu}}, \bibinfo
  {author} {\bibfnamefont {A.}~\bibnamefont {Carvalho}}, \bibinfo {author}
  {\bibfnamefont {Y.}~\bibnamefont {Bao}}, \bibinfo {author} {\bibfnamefont
  {H.}~\bibnamefont {Xu}}, \bibinfo {author} {\bibfnamefont {S.~J.}\
  \bibnamefont {Tan}}, \bibinfo {author} {\bibfnamefont {W.}~\bibnamefont
  {Liu}}, \bibinfo {author} {\bibfnamefont {A.}~\bibnamefont {Castro~Neto}},
  \bibinfo {author} {\bibfnamefont {K.~P.}\ \bibnamefont {Loh}}, \ and\
  \bibinfo {author} {\bibfnamefont {J.}~\bibnamefont {Lu}},\ }\href {\doibase
  https://doi.org/10.1021/acs.nanolett.6b05381} {\bibfield  {journal} {\bibinfo
   {journal} {Nano Lett.}\ }\textbf {\bibinfo {volume} {17}},\ \bibinfo {pages}
  {1970} (\bibinfo {year} {2017})}\BibitemShut {NoStop}%
\bibitem [{\citenamefont {Niu}\ \emph {et~al.}(2020)\citenamefont {Niu},
  \citenamefont {Qiu}, \citenamefont {Wang}, \citenamefont {Zhang},
  \citenamefont {Si}, \citenamefont {Wu},\ and\ \citenamefont
  {Ye}}]{Te_WAL_PRB}%
  \BibitemOpen
  \bibfield  {author} {\bibinfo {author} {\bibfnamefont {C.}~\bibnamefont
  {Niu}}, \bibinfo {author} {\bibfnamefont {G.}~\bibnamefont {Qiu}}, \bibinfo
  {author} {\bibfnamefont {Y.}~\bibnamefont {Wang}}, \bibinfo {author}
  {\bibfnamefont {Z.}~\bibnamefont {Zhang}}, \bibinfo {author} {\bibfnamefont
  {M.}~\bibnamefont {Si}}, \bibinfo {author} {\bibfnamefont {W.}~\bibnamefont
  {Wu}}, \ and\ \bibinfo {author} {\bibfnamefont {P.~D.}\ \bibnamefont {Ye}},\
  }\href {\doibase https://doi.org/10.1103/PhysRevB.101.205414} {\bibfield
  {journal} {\bibinfo  {journal} {Phys.\ Rev.\ B}\ }\textbf {\bibinfo {volume}
  {101}},\ \bibinfo {pages} {205414} (\bibinfo {year} {2020})}\BibitemShut
  {NoStop}%
\end{thebibliography}
\end{document}